\documentstyle[pre,aps,preprint,epsf]{revtex}
\begin{document}
\tolerance=10000
\draft

\title{Universal order-parameter profiles for critical adsorption and
the extraordinary transition: a comparison of $\epsilon$ expansion and Monte
Carlo results\thanks{Paper
presented at the discussion meeting {\it Phase transitions at
Interfaces}, Sept.\ 22--24, 1993, Bad Herrenalb, Federal Republic of
Germany }} \date{\today}
\author{M.\ Smock and H.~W.~Diehl}
\address{%
Fachbereich Physik, Universit\"at - Gesamthochschule - Essen,\\
D-45117 Essen, Federal Republic of Germany }
\author{D.\ P.\ Landau}
\address{%
Center for Simulational Physics, University of Georgia,
Athens, Georgia 30602}
\maketitle
\begin{abstract}
The universal, scaled order parameter profiles $P_{\pm}(z/\xi)$
for critical adsorption of a fluid or fluid mixture onto a
wall or interface, and for the extraordinary transition of the
semi-infinite Ising model, are discussed theoretically, where
$z$ is the distance from the interface, $\xi (T)$ is the
bulk correlation length, and the subscript $+$ ($-$)
refers to the approach from above (below) $T_c$.
Recent results to first order in the
$\epsilon=4-d$ expansion are extrapolated to $d=3$ space dimensions
and compared with new Monte Carlo results. In order to obtain meaningful
extrapolations it is crucial that both the exponential decay
at large $\zeta$ as well as the known
algebraic behavior  $P_{\pm}(\zeta)\sim\zeta^{-\beta/\nu}$ at small $\zeta$
be correctly reproduced. To this end a recently developed
novel RG scheme involving a $z$ dependent amplitude renormalization
is used. Reasonable agreement of our extrapolations with the
Monte Carlo results and some experimental results is obtained.
\end{abstract}

\section*{1.\ Introduction}

The phenomenon of critical adsorption of fluids or binary mixed fluids onto
walls or interfaces has attracted considerable attention, both theoretically
and experimentally, in recent years\cite{ca,FdG,Beysens,Liu,Law}. It was
originally predicted by Fisher and de
Gennes \cite{FdG}. Considering a
semi-infinite fluid bounded by a planar wall at $z=0$ and invoking scaling
ideas, these authors concluded that the
phenomenon has two important signatures:

(i) Between a microscopic scale $a$ and the macroscopic bulk
correlation length $\xi$ there exists a regime
$a\lesssim z\lesssim \xi$ of distances $z$ from the interface
in which the order parameter $m(z)\equiv\langle\phi({\bf x})\rangle$
at position ${\bf x}=({\bf x}_{\parallel},z)$ decays only
{\it algebraically} $\sim z^{-\beta/\nu}$.

(ii) The excess order parameter
\begin{equation}
m_s=\int_0^{\infty}[m(z)-m(\infty)]\,dz\;,
\end{equation}
i.e., the total amount of adsorbed fluid, behaves as
$m_s\sim \tau^{-(\nu-\beta )}$ as $\tau\equiv (T-T_c)/T_c\to 0$,
where $T_c$ is the bulk critical temperature. Since
$\nu-\beta\simeq 0.3$ in three dimensions, $m_s$ {\it diverges}.

On a {\it qualitative} level both signatures (i) and (ii) have
been verified experimentally. However, with regard to detailed
{\it quantitative} investigation of the phenomenon and its
{\it quantitative} comparison with theoretical predictions,
much remains to be done. A detailed theoretical analysis
of optical data for critical adsorption has been made by Liu
and Fisher \cite{Liu}, who found evidence both for (i) as well as for
crossover of $m(z,\tau\ge 0)$ from a power-law regime at small
$\zeta\equiv z/\xi$ to exponential behavior, $m(z)\sim e^{-\zeta}$,
at large $\zeta$. Although they were able to extract also some quantitative
information about the nature of the order parameter profile from the data, the
precision and reliability of their estimates was somewhat limited,
partly because of the limited accuracy of the available experimental
data, but partly also because of the lack of quantitatively reliable
theoretical results needed as input for the analysis of the experiments.
In particular, it became clear that detailed quantitative results
for the scaling functions $P_{\pm}(\zeta)$ governing the asymptotic
form
\begin{equation}
m(z)\approx M_-\,|\tau|^{\beta}\,P_{\pm}(z/\xi)
\end{equation}
for $\tau\to 0^{\pm}$ would be highly desirable. Here $M_-$ is a nonuniversal
constant, which we fix such that $P_-(\infty)=1$. The functions
$P_{\pm}$ also depend on the precise definition of the correlation
length $\xi$, i.e., on the choice of the nonuniversal coefficients
$\xi_0^{\pm}$ in the asymptotic expression
$\xi (\tau\to 0^{\pm})\approx \xi_0^{\pm}\,|\tau |^{-\nu}$. We fix these
by taking $\xi$ to be the {\it true} correlation length, defined through the
requirement that the bulk two-point cumulant function decays
$\propto e^{-|{\bf x}-{\bf x}'|/\xi}$ as
$|{\bf x}-{\bf x}'|\to\infty$.

Recently two of us \cite{Smock,Diehlbuns} succeeded in computing the first two
terms in the $\epsilon =4-d$ expansion of $P_{\pm}(z/\xi)$. Upon
extrapolation to $d=3$ space dimensions explicit results for
$P_{\pm}(\zeta)$ were obtained, which had the correct physical
features, displaying, in particular, a smooth and monotonous crossover from the
behavior $P_{\pm}(\zeta)\sim \zeta^{-\beta/\nu}$ as $\zeta\to 0$ to the
exponential decay at large $\zeta$. In the present paper we elaborate on these
results, extending them and checking them versus Monte Carlo results. We begin
with a brief summary of
the $\epsilon $-expansion results of Ref.\  \onlinecite{Smock}, recalling that
their naive extrapolation to $d=3$ would yield profiles $P_{\pm}(\zeta)$ with a
totally unacceptable, incorrect short-distance behavior. We show that this
problem can be overcome in a systematic manner by means of a recently developed
novel renormalization group (RG) scheme \cite{SD}, which ensures the proper,
exponentiated form of the leading short-distance singularity. We then
compare the so-obtained extrapolations $P_{\pm}(\zeta, d=3)$ with the results
of Monte Carlo calculations. Finally, we present the
$\epsilon$ expansion of a number of universal amplitude
ratios, estimate their values for $d=3$, and
compare them with experimental estimates, if possible.

\section*{2.\ Results of the $\epsilon$ expansion and beyond}

The analytic results of Ref.\ \onlinecite{Smock} were obtained by means of RG
improved perturbation theory applied to the
semi-infinite $\phi^4$ model \cite{Diehl}. Since
this model and its physics is reviewed in a separate contribution by Diehl
\cite{Diehlbuns} in these proceedings, we will focus on the explanation of the
results. Critical adsorption (or the so-called {\it normal} transition) is
described by a fixed point ${\cal P}_{\text{ca}}^*$; the extraordinary
transition by a fixed point ${\cal P}_{\text{ex}}^*$. Although being
located in
different regions of parameter space of the model, these fixed points are
equivalent in the sense that they yield identical results for the scaling
functions $P_{\pm}$ and other quantities to arbitrary order of the $\epsilon$
expansion. In Ref.\ \onlinecite{Smock} the first two terms in the expansion
\begin{equation}
P_{\pm}(\zeta ;\epsilon)=P_{\pm}(\zeta ;\epsilon=0)+\epsilon\,
\partial_{\epsilon} \,P_{\pm}(\zeta ;\epsilon=0)+O(\epsilon^2) \label{epsex}
\end{equation}
were obtained. Note that in contrast to our convention here the second-moment
definition of $\xi$ was used in Ref.\ \onlinecite{Smock}.
We denote this latter correlation length
as $\hat{\xi}$, reserving the symbol $\xi$ exclusively for the true
correlation length. The difference between $\xi$ and $\hat{\xi}$ is
fairly small near criticality since the associated amplitudes
$\xi_0^{\pm}$ and $\hat{\xi}_0^{\pm}$ differ by less
than a few per cent \cite{Tarko}. The
scaling functions $P_{\pm}$ and $\hat{P}_{\pm}$ corresponding to these two
conventions can be easily transformed into each other using
\begin{equation}
\hat{P}_{\pm}(\zeta )=P_{\pm}(\zeta\,\xi_0^{\pm}/\hat{\xi}_0^{\pm})\;.
\end{equation}
Since $\xi_0^{+}/\hat{\xi}_0^{+}=1+O(\epsilon^2)$
and $\xi_0^{-}/\hat{\xi}_0^{-}=1+(\epsilon/12)(11/2 -\pi
\sqrt{3})+O(\epsilon^2)$ \cite{Brezin}, the
$\epsilon$ expansions of $P_+$ and $\hat{P}_+$ agree to
order $\epsilon$, but those of $P_-$ and $\hat{P}_-$ differ
already at order $\epsilon$.

As $\zeta\to 0$,
the two terms in (\ref{epsex}) behave as
$P_{\pm}(\zeta ;\epsilon=0)\approx c_{\pm}\,\zeta^{-1}$ and
$\partial_{\epsilon} \,P_{\pm}(\zeta ;\epsilon=0)\approx (c_{\pm}\,
 /6 \zeta)\ln\zeta$.
If one extrapolated this result naively to $d=3$
by setting $\epsilon =1$, one would obtain a totally unacceptable
short-distance behavior of the form
$P_{\pm}(\zeta)\sim [1+(\epsilon /6)\ln\zeta ]/\zeta $. As discussed in
Refs.~\onlinecite{Smock} and \onlinecite{Diehlbuns}, the correct limiting
form is
\begin{equation}
P_{\pm}(\zeta\to 0) \approx
\zeta^{-\beta/\nu}\big(c_{\pm}+a_{\pm}\,\zeta^{1/\nu}+a'_{\pm}\,\zeta^{2/\nu}
+b_{\pm}\,\zeta^d+\,... \big)\;.
\label{asfP}
\end{equation}
Equation (\ref{epsex}) is consistent with
(\ref{asfP}) in that the $\epsilon$ expansion of the latter agrees with the
limiting form of (\ref{epsex}) for small $\zeta$. The coefficients
have the expansions
\begin{equation}
a_+= \sqrt{2} \,\left(-\case{1}/{6}+ \case{\epsilon}/{216} \,
(1-6\, C_E-6 \,\ln\! 2) \right)+O(\epsilon^2) \;, \label{a+}
\end{equation}
\begin{equation}
a_-= \case{1}/{6} +\case{\epsilon}/{216} \,\left(
6 C_E+8- \pi \, 3 \sqrt{3}  \right) +O(\epsilon^2)
\;, \label{a-} \end{equation}
\begin{equation}
a_+'=\case{ \sqrt{2}}/{36}+O(\epsilon)\;,\qquad a_-'=
-\case{1}/{72} +O(\epsilon) \;,\label{ap+}
\end{equation}
\begin{equation}
b_+= - \case{ \sqrt{2}}/{120}+O(\epsilon) \;,\qquad b_-=
 -\case{1}/{60} +O(\epsilon)\;, \label{b+}
\end{equation}
\begin{equation}
c_+=\sqrt{2} \, \left[ 1+ \case{\epsilon}/{12} \, (6 \, C_E+2 \, \ln\!
2 - 13)
\right] +O(\epsilon^2) \;, \label{c+}
\end{equation}
\begin{equation}
c_-= 2+\case{\epsilon}/{6}  \, \left( 6 \, C_E-16 +\pi \sqrt{3} \right) +O(\epsilon^2)
\;,
\label{c-}
\end{equation}
where $C_E=0.5772\ldots$ is Euler's constant. Due to the difference between
$\xi$ and $\hat{\xi}$, the $O(\epsilon )$ terms of $a_-$ and $c_-$ differ
from those of $\hat{a}_-$ and $\hat{c}_-$ given in Ref.\ \onlinecite{Smock}.

For large arguments the scaling functions have the
expected exponentially decaying forms
\begin{equation}
P_+(\zeta\to\infty) \approx
2 \sqrt{2} \, \left\{ 1 +\epsilon\left[ \case{1}/{4}+
\case{1}/{6}\,\ln 2
 -\case{2}/{3}	\pi  \left( 1-2 \sqrt{3}\right)
+\case{ 4 \sqrt{3}}/{3}
 \ln \case{2 \sqrt{3}+3}/{2 \sqrt{3} -3 }  \right]+
O(\epsilon^2) \right\}\,
e^{-\zeta} \label{P+as}
\end{equation}
and
\begin{equation}
P_-(\zeta\to\infty )\approx 1+ \left[ 2 -\epsilon \,
\case{\pi}/{6} \left( 4- 1/\sqrt{3}
\right) \right] +O(\epsilon^2)] \,e^{-\zeta}
\; ,
\label{P-as}
\end{equation}
where it should be noted that $e^{-\zeta}$ would get
replaced by $e ^{-\text{const }\zeta}$ for any choice of the
correlation length other than the true one.

In light of the discussion given in Ref. \cite{Liu} we also determined
the next-to-leading terms in the asymptotic expansion
\begin{equation}
P_+ (\zeta) = P^{(\infty )}_{+,1} \,e^{- \zeta} + P^{(\infty )}_{+,2}
\,e^{-2 \zeta}
+P^{(\infty )}_{+,3} \,e^{-3 \zeta} + \ldots\;.
\end{equation}
The value of $P_{+,1}^{\infty}$ can be read off of (\ref{P+as}).
For the remaining coefficients one has
\begin{equation}
 P^{(\infty )}_{+,2}=O(\epsilon^2) \,
\end{equation}
and
\begin{equation}
P^{(\infty )}_{+,3}=
 1 +\epsilon\left[\case{1}/{4}+ \case{1}/{6}\ln 2
 -2\pi \, ( 1-2 \sqrt{3})-4 \sqrt{3}
 \ln \case{2 \sqrt{3}+3}/{2 \sqrt{3} -3}\right]
+O(\epsilon^2) \, .
\end{equation}

The above results provide valuable quantitative information about the nature
of the scaling functions beyond what is known from scaling considerations.
The limiting forms of $P_{\pm}(\zeta )$ at large and small $\zeta$ can be
extrapolated to $d=3$ in a straightforward manner either by simply setting
$\epsilon =1$ in the corresponding coefficients or, preferably, by using
improved extrapolation techniques (e.g., Pad\'e analyses, into which the exact
results for $d=2$ have been incorporated \cite{Gero}). To obtain extrapolations
$P_{\pm}(\zeta, d=3)$ giving a proper description
of the $\zeta$ dependence for
all values of $\zeta$ is a more challenging problem. A prime requirement is
that both the leading asymptotic terms at small {\it and} large $\zeta$ be
correctly reproduced. As we have seen above, the most naive extrapolation
procedure --- setting $\epsilon=1$ in (\ref{epsex}) --- grossly fails in this
regard. A number of alternative extrapolation schemes have been suggested
\cite{PL,Smock}, but all of these contain some degree of arbitrariness. Only
recently a more systematic way of handling this problem was developed
\cite{SD}.

The crux of this method is a specially adapted, novel RG scheme. Its details
are beyond the scope of the present paper, so we restrict ourselves
to a few remarks. In renormalized field theory the amplitude of the order
parameter $\phi({\bf x})$ usually is reparametrized in a $z$ {\it independent}
fashion to define the renormalized density $\phi_R=Z_{\phi}^{-1/2}\,\phi $.
Here the amplitude renormalization factor $Z_{\phi}(u)$ depends
on $u$, the dimensionless renormalized coupling constant, but not on $z$. The
basic new element of the approach of Ref.\ \cite{SD} is that in place of
$Z_{\phi}$ a $z$ {\it dependent} amplitude renormalization factor
$\tilde{Z}(u,z\mu)$ is used, where $\mu$ is an arbitrary reference
momentum. This generalized renormalization factor $\tilde{Z}$ is required to
absorb in addition to the usual ultraviolet singularities (corresponding to
poles in $\epsilon$ in the dimensionally regularized theory) also the
short-distance singularities $\sim z^{-1}\ln (\mu z)$. That is, we require that
the renormalized profile $m_R(z,u,\tau;\mu)=\tilde{Z}^{-1/2}\,m(z)$ have
a finite limit
\begin{equation}
\lim_{z\to 0}z\,m_R(z,u,\tau;\mu)\;.
\end{equation}
At one-loop order a convenient choice is
\begin{equation}
\tilde{Z}=1+\case{3}/{2}\,u\ln\!\left(1+\case{1}/{\mu z}\right)+O(u^2)\;.
\end{equation}

The advantage of this scheme becomes clear when the resulting RG equations are
exploited. Due to the $z$ dependence of $\tilde{Z}$, these are somewhat more
complicated than usual, albeit solvable by standard methods. The resulting
RG-improved perturbation theory yields scaling functions
$P_{\pm}(\zeta,\epsilon)$ which exhibit for all $\epsilon>0$ a smooth
crossover from the correct short-distance form $\sim \zeta^{-\beta/\nu}$ to the
exponential decay at large $\zeta$. Accordingly, meaningful extrapolations are
obtained upon setting $\epsilon =1$ (see Figs.\ 1 and 2).

\section*{3.\ Monte Carlo Results}

Monte Carlo calculations have proved to be a very powerful alternative tool
for studying critical phenomena in confined geometries.
To get independent information on the scaling functions $P_{\pm}$ we
analyzed data of a Monte Carlo simulation of an
$ L \times L \times D $ $ (L=128, D=80)$  simple-cubic	Ising film with
periodic boundary conditions in the $L$-directions and free
boundaries in the $D$-direction. The spins were assumed to interact via
ferromagnetic nearest-neighbor exchange couplings that take the values $J_1$
and $J$ for
all bonds lying entirely within the two boundary layers and all other bonds,
respectively. All bulk and surface magnetic fields were set to zero. To
observe the universality class of the extraordinary transition,
the surface coupling
$J_1$ was chosen to be supercritical, i.e., larger than the critical value
for the occurence of a special transition in the thermodynamic limit
$L,\,D\to\infty$ (cf.\ Refs.\ \onlinecite{Landau,Diehlbuns,Diehl}).

To model critical adsorption of fluids it would seem more realistic to
take $J_1$ subcritical and to add a surface magnetic field $h_1>0$
(accounting for the wall-fluid interaction), i.e., to consider the
so-called {\it normal} surface transition {\cite{Diehlbuns} of the model.
However, as discussed in
Ref.\ \onlinecite{Diehlbuns} and demonstrated exactly in Ref.\ \onlinecite{BD},
the surface critical behavior at critical adsorption and the normal
transition is representative of the same universality class as the
extraordinary transition.

The Monte Carlo data show that the behavior of the order parameter $m(z)$
for $z\ll\xi$ is well described by $m(z)\propto (z+z_e)^{-\beta/ \nu} $,
where the exponent is in conformity with the value
$\beta/ \nu = 0.519 \pm 0.007$ quoted in Ref.\ \onlinecite{Liu}. The quantity
$z_e$ is a microscopic (`extrapolation') length depending on $J_1/J$,
which ensures that $m_1=m(0)$ is finite. It drops out in the scaling regime
$z\gg z_e$. In accordance with our expectations we found it to have the same
value for $T>T_c$ and $T<T_c$. For large distances $z$, which are still
sufficiently small so that finite size effects are negligible, the profiles
$m_{\pm}(z)$ decay exponentially on the scale of $\xi$, where
the value of $\xi$ is in conformity with the results of Ref.\
\onlinecite{Tarko}. At even larger values of $z$
finite size effects are clearly visible.

In order to check whether scaling holds we analyzed the data for $m(z)$
in the first twenty layers closest to the surface, using the scaling ansatz
\begin{equation}
m(z) \simeq {M_-} |\tau |^{\beta } P_{\pm} ( \case{z+z_e}/\xi)
\label{sc}
\end{equation}
for reduced temperatures $\tau$ in the ranges given in Figs.\ 1 and 2. As is
borne out by these figures, the data collapsing on a single curve $P_{\pm}$
works reasonably well both for $\tau>0$ and $\tau<0$, within the
limitations caused by statistical error and finite size effects.
The 
resulting Monte Carlo profiles $P_{\pm}(\zeta)$ agree qualitatively  
quite well with our extrapolated $\epsilon$ expansion
results. Quantitatively, the agreement is not very satisfactory. 
To appreciate these results, one should note, however, that our 
extrapolated RG results are {\it free of
any adjustable parameter.} Likewise, no adjustable parameter remains in the 
Monte Carlo results for $P_{\pm}$, since the scales and amplitudes are fixed  
by the required large-distance forms (\ref{P+as}) and (\ref{P-as}). 
(One probably could get improved agreement
at intermediate values of $\zeta$ by exploiting
the freedom to make $O(\epsilon^2)$ errors at order $\epsilon$,
but in contrast to our procedure, this would be unsystematic.)

The results for $P_{\pm}$ may be used to estimate various universal ratios.
The ratio
\begin{equation}
R_{MA}\equiv \int_0^{\infty}d\zeta\,P_+(\zeta )/
\int_0^{\infty}d\zeta\,[P_-(\zeta )-1]\;,
\end{equation}
which is proportional to the ratio of the amplitudes of the excess order
parameter $m_s$ above and below $T_c$, has been investigated
experimentally by Law and Smith \cite{Law} recently. Integration of our
Monte Carlo results yields $R_{MA}\simeq 1.14$ in good agreement with their
experimental estimate $1.18\pm 0.13$. Integration of our extrapolated
RG results yields a somewhat higher value $\simeq 1.3$. Of considerable
interest are also the universal ratios $c_{\pm}$. From our Monte Carlo
results we find the estimates $c_+=0.866\pm 0.07$ and $c_-=1.22\pm 0.08$.
Fl\" oter and Dietrich \cite{Gero} estimated $c_+$ theoretically
by means of  Pad\'e approximants into which
both the $\epsilon$ expansion results of Ref.\ \onlinecite{Smock} and the
exact results for $d=2$ were incorporated. They also extracted values
of $c_+$ from a number of experiments. These experimental estimates were
found to scatter around their theoretical estimate $c_+\simeq 0.95$, which is
in accordance with our Monte Carlo estimate. Finally, we mention that the
universal ratio $P_{+,1}^{\infty}/c_+$ was considered first by Liu and
Fisher \cite{Liu}. From their analysis of critical adsorption data they
obtained the estimate $\simeq 0.85$. Our Monte Carlo data yield a    
considerably larger value of $\simeq 1.85$. This discrepancy might be due to the
limited accuracy of the experimental data available at that time. A problem
could also be that the approximate forms of $P_+(\zeta)$ used in the analysis
of the experimental data were not accurate enough.
(As discussed by Liu and Fisher  \cite{Liu} the estimate is rather sensitive to the chosen form of $ P_+$.)
\section*{4.\ Concluding remarks}

In summary, we presented results for the universal scaled order parameter
profiles $P_{\pm}(\zeta)$ at critical adsorption and the extraordinary
transition. These were obtained by two distinct methods, namely, by RG-improved
perturbation theory in $4-\epsilon$ dimensions and by Monte Carlo simulations.
To obtain meaningful extrapolations of the RG results to the physically
interesting dimension $d=3$, a novel renormalization scheme was used that
guarantees the correct exponentiation of the leading short-distance
singularity. Both the latter method and the Monte Carlo simulations
yield profiles for $d=3$, which exhibit the expected smooth
and monotonous crossover from the power-law behavior (\ref{P+as}) at small
$\zeta$ to the large-distance form (\ref{P+as}). The results of both methods
are in fair agreement with each other. The
knowledge of the profiles can be utilized to estimate various universal
amplitude
ratios, which can also be determined by experiments. The agreement of our
theoretical estimates with the few
available experimental estimates \cite{Law,Gero} of the universal numbers $c_+$
and $R_{MA}$ is encouraging. Yet, much more detailed
experimental checks of the theoretical predictions are clearly needed.

\section*{Acknowlegements}

We would like to thank G.\ Fl\"oter and S.\ Dietrich for keeping us informed
about their work and for helpful discussions. This work has been supported
in part by Deutsche Forschungsgemeinschaft through Sonderforschungsbereich 237.

\figure{Fig.\ 1: \  Scaling function $ P_+ (\zeta ) $. The dashed curve 
shows the mean-field approximation. The full curve is the extrapolation 
to $d=3$ of the one-loop RG result obtained with the help of our specially 
adapted renormalization scheme. The symbols show   
Monte Carlo results for different 
reduced temperatures.

\figure{Fig.\ 2: \  Scaling function $ P_- (\zeta ) $. The dashed and full 
curves are the $\tau<0$ analogs of the curves in Fig.\ 1.  The symbols 
represent Monte Carlo results for different reduced temperatures.

\newpage

\begin{figure}[h]
\epsfxsize=16cm
\[\epsffile{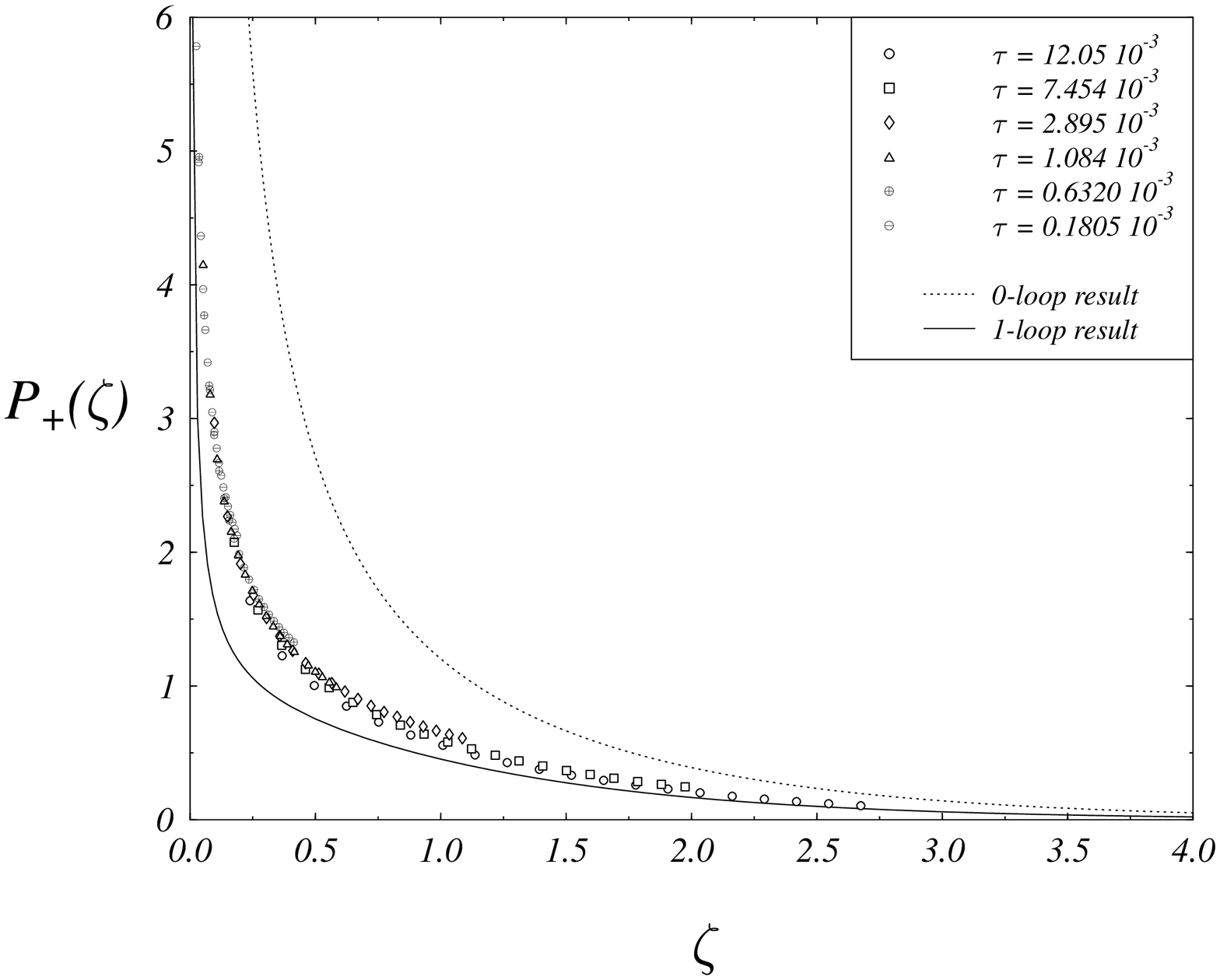}\]
\end{figure}

\figure{Fig.\ 1: \  Scaling function $ P_+ (\zeta ) $. The dashed curve 
shows the mean-field approximation. The full curve is the extrapolation 
to $d=3$ of the one-loop RG result obtained with the help of our specially 
adapted renormalization scheme. The symbols show   
Monte Carlo results for different 
reduced temperatures.

\begin{figure}[h]
\epsfxsize=16cm
\[\epsffile{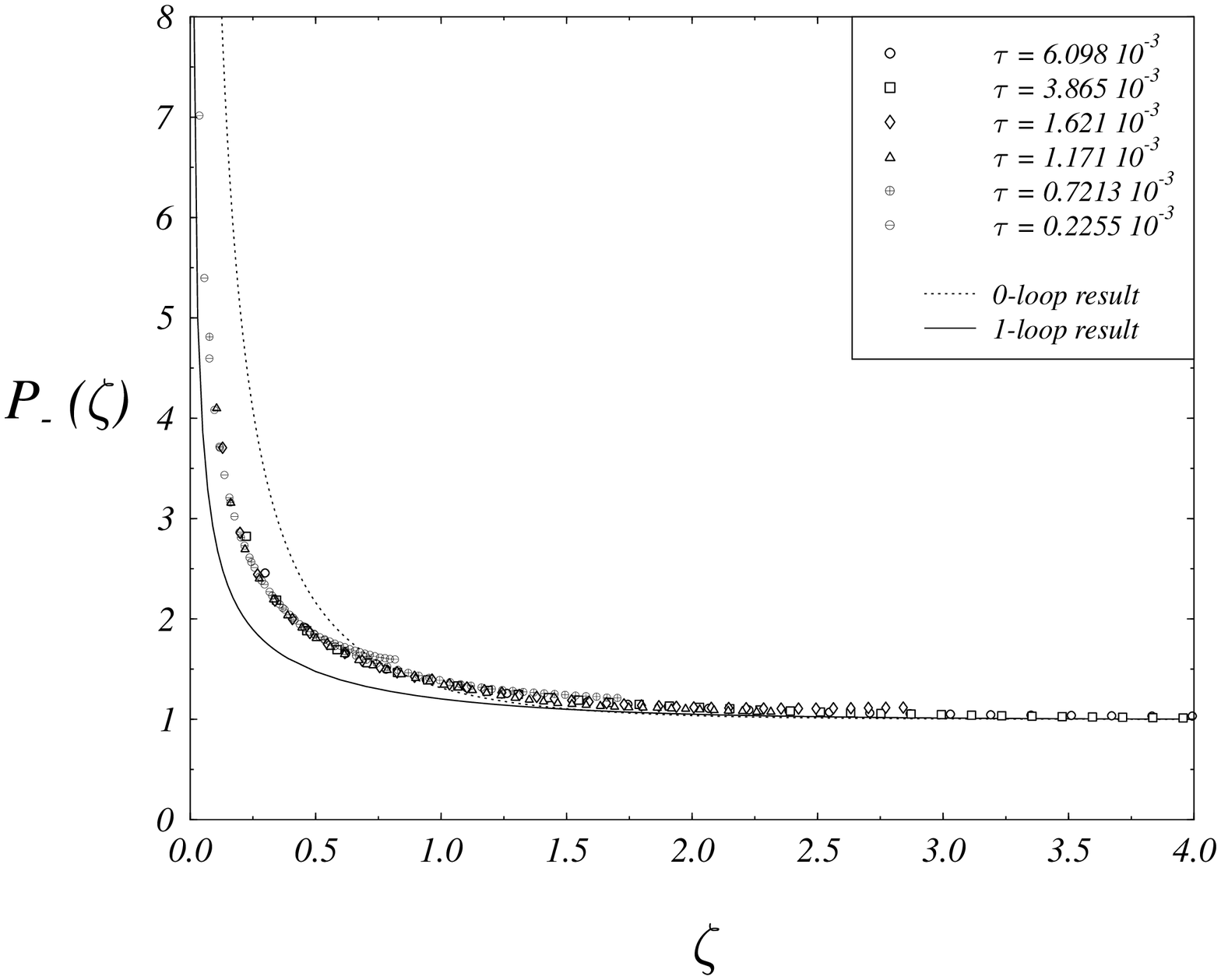}\]
\end{figure}

\figure{Fig.\ 2: \  Scaling function $ P_- (\zeta ) $. The dashed and full 
curves are the $\tau<0$ analogs of the curves in Fig.\ 1.  The symbols 
represent Monte Carlo results for different reduced temperatures.


\begin{references}
\bibitem{ca}{This phenomenon was originally predicted in Ref.\
\onlinecite{FdG} and first observed in Ref.\ \onlinecite{Beysens}.
For a recent discussion of pertinent experimental work and a
list of references, see Ref.\ \onlinecite{Liu} and Ref.\ \onlinecite{Law}}.
\bibitem{FdG}{M.\ E.\ Fisher and P.-G.\ de Gennes, C.\ R.\ Acad.\ Sci.\ B
{\bf 287}, 207 (1978)}
\bibitem{Beysens}{D.\ Beysens and S.\ Leibler,
J.\ Phys.\ Lettres {\bf 43}, L 133 (1982).}
\bibitem{Liu}{A.\ J.\ Liu and M.\ E.\ Fisher, Phys.\ Rev.\ A {\bf 40}, 7202
(1989).}
\bibitem{Law}{B.\ M.\ Law, paper in these proceedings.}
\bibitem{Smock}{H.\ W.\ Diehl  and M.\ Smock, Phys.\ Rev.\ B {\bf 47},
5841 (1993); Erratum Phys.\ Rev.\ B {\bf 48}, 6470 (1993).}
\bibitem{Diehlbuns}{H.\ W.\ Diehl, accompanying paper}
\bibitem{SD}{M.\ Smock and H.\ W.\ Diehl, to be published.}
\bibitem{Diehl}{For background and references, see H.\ W.\ Diehl, in {\it Phase
Transitions and Critical Phenomena},
edited by C.\ Domb and J.\ L.\ Lebowitz
(Academic Press, London, 1986), Vol.\
X, p.\ 75.}
\bibitem{Tarko}{H.\ B.\ Tarko and M.\ E.\ Fisher,
Phys.\ Rev.\ B {\bf 11}, 1217 (1975).}
\bibitem{Brezin}{E.\ Br\'ezin, J-C.\ Le\ Guillou and J.\ Zinn-Justin,
Phys.\ Lett.\ {\bf 47 A}, 285 (1974).}
\bibitem{Gero}{G.\ Fl\"oter and S.\ Dietrich, preprint
BUGH Wuppertal.}
\bibitem{PL}{L.\ Peliti and S.\ Leibler, J.\ Phys.\ C {\bf 16}, 2635 (1983).}
\bibitem{Landau}{ For details on the algorithm, see
 D.\ P.\ Landau and K.\ Binder, Phys.\ Rev.\ B {\bf 41}, 4786
(1990).}
\bibitem{BD}{T.\ W.\ Burkhardt and H.\ W.\ Diehl, to be published.}
\bibitem{Law}{B.\ M.\ Law, Phys.\ Rev.\ Lett.\ {\bf 67}, 1555 (1991); D.\ S.\
P.\ Smith and B.\ M.\ Law, J.\ Chem.\ Phys., Dec.\ 15 (1993);
B.\ M.\ Law, this volume.}
\end{references}
\end{document}